\documentclass{ws-mplb}

\newcommand{\REVIEW}[4]{{#1} \textbf{#2}, #3 (#4)}

\begin{document}

\markboth{Korshunov, Ovchinnikov, Shneyder, Gavrichkov, Orlov, Nekrasov, Pchelkina}{Cuprates, Manganites, and Cobaltites: Multielectron Approach to the Band Structure}

%
\catchline{}{}{}{}{}
%

\title{CUPRATES, MANGANITES, AND COBALTITES: MULTIELECTRON APPROACH TO THE BAND STRUCTURE}

\author{\footnotesize M.M. KORSHUNOV$^{1,2}$, S.G. OVCHINNIKOV$^{1,2}$, E.I. SHNEYDER$^{1,3}$, V.A. GAVRICHKOV$^{1,2}$, Yu.S. ORLOV$^{1,2}$, I.A. NEKRASOV$^{4}$, Z.V. PCHELKINA$^{5}$}

\address{$^{1}$L.V. Kirensky Institute of Physics, Siberian Branch of Russian Academy of Sciences, Krasnoyarsk 660036, Russia\\
$^{2}$Siberian Federal University, Svobodny Prospect 79, Krasnoyarsk 660041, Russia\\
$^{3}$ Reshetnev Siberian State Aerospace University, Krasnoyarsk 660014, Russia\\
$^{4}$ Institute for Electrophysics, Russian Academy of Sciences,
Ekaterinburg 620016, Russia\\
$^{5}$ Institute for Metal Physics, Russian Academy of Sciences, Ekaterinburg 620219, Russia\\
sgo@iph.krasn.ru}

\maketitle

\begin{history}
\received{(Day Month Year)}
\revised{(Day Month Year)}
\end{history}

\begin{abstract}
High-$T_c$ superconductors with CuO$_2$ layers, manganites La$_{1-x}$Sr$_x$MnO$_3$, and cobaltites LaCoO$_3$ present several mysteries in their physical properties. Most of them are believed to come from the strongly-correlated nature of these materials. From the theoretical viewpoint, there are many hidden rocks in making the consistent description of the band structure and low-energy physics starting from the Fermi-liquid approach. Here we discuss the alternative method -- multielectron approach to the electronic structure calculations for the Mott insulators -- called LDA+GTB (local density approximation + generalized tight-binding) method. Its origin is a straightforward generalization of the Hubbard perturbation theory in the atomic limit and the multiband $p-d$ Hamiltonian with the parameters calculated within LDA. We briefly discuss the method and focus on its applications to cuprates, manganites, and cobaltites.
\end{abstract}


\section{Introduction}
\label{Int_LDA_GTB:1}

Fermi-liquid approach is one of the most attractive and useful in the modern condensed matter physics. Its cornerstone is a quasiparticle defined as a renormalized electron -- the one surrounded by the cloud of excitations. Such concept allows to use a number of the well-developed theoretical tools like diagram technique, path integrals, etc. The strong point of the Fermi-liquid approach is generality but in many cases it becomes a weak point. When it comes to real material-specific questions, the starting point for the study should depend on the particular material's characteristics. Huge leap in this direction was made in mid 60th by Hohenberg, Kohn, and Sham\cite{Hohenberg_1964,Kohn_1965} who formulated a density functional theory (DFT). Because its starting point is the Shr\"odinger equation for the particular atomic arrangement, orbital, and spin configurations, this theory is often referred to as the `first principles' or the `\textit{ab initio}' calculations. Augmented with the local density approximation (LDA) or the generalized gradient approximation (GGA) for the initially unknown quantity, exchange-correlation energy, DFT provides quantitative description of the ground state energy and the band structure of various atoms, molecules, and crystalline solids, especially containing $s$ and $p$ atoms (see e.g.\cite{Jones_1989}).

There are many cases, however, when the basic concept of the Fermi-liquid brakes and the system becomes a non-Fermi-liquid. Examples include temperature-dependent metal-insulator transitions in cobaltites and a pseudogap state in cuprates. In this case, one-electron approaches like LDA and GGA give qualitatively wrong results. In particular, LDA failed to describe transition metal oxides with partially filled $3d$-orbitals. The most pronounced failure is that LDA predicts La$_2$CuO$_4$ to be a metal whereas experimentally it is an insulator. The root of the problem is the unscreened on-site Coulomb interaction (Hubbard repulsion)\cite{r7LDAGTB}. In a single-band system on the mean-field level if Hubbard repulsion $U$ is larger than the bandwidth $W$, it splits the band into two Hubbard subbands with a gap $\propto U$. Spectral weight of a quasiparticle is redistributed between these subbands. At a half-filling, the Fermi level is inside the gap and the system is an insulator. In a multiorbital system, along with the Hubbard repulsion other local interactions like the Hund's exchange $J_H$ and the interorbital Hubbard repulsion $U'$ are present and provide a rich set of physical properties. Opening of the Hubbard gap and moreover the major role played by the local interactions near the half-filling are beyond the scope of LDA and GGA.

There are several extensions to LDA which includes or simulates the effects of the on-site interactions. One of them is LDA+U\cite{Anisimov_1991} and another one is SIC-LSDA (self-interaction--corrected local spin density approximation)\cite{Svane_1990}. Both methods consider local interactions in the Hartree-Fock sense and result in the antiferromagnetic insulator as the ground state for La$_2$CuO$_4$ contrary to LDA, but the origin of the insulating gap is incorrect. In both LDA+U and SIC-LSDA, it is formed by the local single-electron states splitted by the spin or orbital polarization. Therefore, the paramagnetic phase above the N\'eel temperature $T_N$ of the undoped La$_2$CuO$_4$ will be metallic in spite of strong correlation regime $U \gg W$. There is one more significant drawback in these approximations, namely, they disregard the redistribution of the spectral weight between the Hubbard subbands. Latter effect is incorporated in a different approach to \textit{ab initio} calculations for strongly correlated systems - LDA+DMFT (LDA+dynamical mean field theory)\cite{Anisimov_1997,Lichtenstein_1998,Held_2001,Kotliar_2006}. The method is based on the self-consistent procedure where the LDA band structure is used to calculate the electron self-energy in DMFT. DMFT utilizes the fact that in the infinite dimensional limit of the Hubbard model, $D \to \infty$, the self-energy is momentum independent, $\Sigma(\mathbf{k},\omega) \to \Sigma(\omega)$\cite{Metzner_1989,Vollhardt_1993,Georges_1996}. The remaining frequency dependence is exact in $D \to \infty$ limit and carries very important information about dynamical correlations and Mott-Hubbard transition. On the other hand, the spatial correlations become crucial in low-dimensional systems like layered high-$T_c$ cuprates. That is why the correct band dispersion and spectral intensities for these systems cannot be obtained within LDA+DMFT. Natural extension of this method, LDA+cluster or cellular DMFT\cite{Hettler_1998,Kotliar_2001,Potthoff_2003,Maier_2005}, and SDFT (spectral density functional theory)\cite{Savrasov_2004} provides momentum dependent self-energy and thus allow for the non-local correlations.

Here we are going to discuss the alternative approach -- the LDA+GTB method -- to study the Mott-Hubbard insulators. From the very beginning the GTB (generalized tight binding) method has been suggested to extend the microscopic band structure calculations to take the strong electronic correlations (SEC) into account in the Mott-Hubbard insulators like the transition metal oxides\cite{r1LDAGTB}. Similar to conventional tight binding (TB) approach we start with a particular local electron states (with all multiorbital effects, symmetry and chemistry) and then by a Fourier transform move to the momentum space and obtain a band structure. Because of SEC we can not use free electron local states, our local fermion in a $d$-orbital system is a quasiparticle given by the excitations between multielectron $d^n$ and $d^{n \pm 1}$ terms contrary to the conventional TB. In other words, GTB is the strongly correlated version of the TB method. The first computer codes and successful application of GTB has been developed for cuprates\cite{r2LDAGTB}. That version used the multiband $p-d$ model for La$_2$CuO$_4$\cite{r3LDAGTB} with a lot of empirical parameters in the Hamiltonian. To ``cook'' the \textit{ab initio} approach, the hybrid LDA+GTB method has been developed\cite{r4LDAGTB}. Afterward, similar ideas have been used to study the GTB band structure of manganites La$_{1 - x}$Sr$_x$MnO$_3$\cite{r5LDAGTB} and cobaltites LaCoO$_3$\cite{r6LDAGTB}.

The LDA+GTB may be considered as the straightforward development of the Hubbard atomic representation approach\cite{r7LDAGTB} to real materials like $3d$ metal oxides\cite{ManciniBook}. Indeed, the GTB is a specific version of cluster perturbation theory (CPT) in the Hubbard $X$-operators representation\cite{r8LDAGTB}.

Later we are going to describe the method and its applications to three classes of materials with SEC: cuprates, manganities and cobaltites. The rest of the Review is organized as follows. In the next Section we provide the main ideas and technical steps of the LDA+GTB and consider different approximations to solve the Dyson equation in the $X$-representation. In Section~\ref{CalcEx:5} we discuss the LDA+GTB band structure of La$_{2 - x}$Sr$_x$CuO$_4$. Section~\ref{CalcEx:6} and Section~\ref{lacoo3:7} are devoted to manganites and cobaltites, respectively. Section~\ref{EndGTBLDA:8} is the conclusion.

\section{LDA+GTB method}
\label{QPdef:2}
The very first step in deriving the GTB method is to define an ``electron'' in system with SEC. Due to the strong interactions the free electrons are so heavily renormalized, so that the new objects are unlikely to be called ``the electrons''. Still, we would like to generate the output of the theory in terms of the Green function that is the one-particle property, so the ``electron Green function''. The goal is to find the electron Green function or, in other words, to define the electron. According to the exact Lehmann representation\cite{r11LDAGTB}, at $T=0$ the electron Green function can be written as
\begin{equation}
\label{QPdef:eq1}
G_\sigma(\vec{k},\omega) = \sum\limits_m \left(\frac{A_m(\vec{k},\omega)} {\omega - \Omega_m^+} + \frac{B_m(\vec{k},\omega)}{\omega - \Omega_m^-} \right).
\end{equation}
where $\Omega_m^+ = E_m (N + 1) - E_0 (N) - \mu $, $\Omega _m^ -   = E_0 (N) - E_m (N - 1) - \mu $, $\mu$ is the chemical potential, and numerators are equal to
\[
\begin{array}{lcr}
A_m \left( {\vec{k},\omega } \right) &=& \left| {\left\langle {0,N}
\right|a_{\vec{k}\sigma } \left| {m,N + 1} \right\rangle } \right|^2, \\
B_m (\vec{k},\omega ) &=& \left| {\left\langle {m,N - 1} \right|a_{\vec{k}\sigma } \left| {0,N} \right\rangle } \right|^2.
\end{array}
\]
Here, $\left| {m,N} \right\rangle $ is the $m$-th eigenstate of the $N$ electron system, $H\left| {m,N} \right\rangle = E_m \left| {m,N} \right\rangle$.

Since each single pole contribution on the right-hand side of Eq.~(\ref{QPdef:eq1}) corresponds to some QP, we interpret the Lehmann representation in the following way: electron is the linear superposition of QPs with the energies $\Omega_m^+$ ($\Omega_m^-$) for electron addition (removal) and with the spectral weights $A_m$ ($B_m$). At finite temperatures, the Lehmann representation determines the QP as the excitation between two arbitrary $\left| {m,N + 1} \right\rangle $  and $\left| {n,N} \right\rangle $ terms with the energy $\Omega _{mn}  = E_m  - E_n $ and a temperature dependent spectral weight\cite{r11LDAGTB}. This definition is very clear. Unfortunately in general case it cannot be used straightforwardly because the exact eigenstates $\left| {m,N} \right\rangle $ and the eigenenergies $E_m$ are unknown. The Landau Fermi-liquid QP is a specific case of the Eq.~(\ref{QPdef:eq1}) with only one QP close to the Fermi level. For the free electron with energy $\varepsilon_0$, all QP energies are equal to $\Omega_m^+ = \Omega_m^- = \varepsilon_0 - \mu$. Now we are going to show that the GTB method is the perturbative realization of the Lehmann representation.

As any other CPT approach, the GTB method starts with the exact diagonalization (ED) of the intracell part ($H_c$) of the multielectron Hamiltonian and treats the intercell part ($H_{cc}$) by a perturbation theory. Thus we make a realization of the Lehmann representation inside one unit cell with all local QP energies and spectral weights calculated via ED. The total LDA+GTB procedure consists of the following steps\cite{r4LDAGTB}:
\begin{description}
  \item[\textbf{Step I: LDA.}] Calculation of the LDA band structure, construction of Wannier functions with the given symmetry, and computation of the one- and two-electron matrix elements of the TB Hamiltonian with the local and nearest-neighbor Coulomb interactions.
  \item[\textbf{Step II: ED.}] Separation of the total Hamiltonian $H$ into the intra- and inter-cell parts, $H=H_c+H_{cc}$, where $H_c$ represents the sum of the orthogonal unit cells, $H_c = \sum\limits_f {H_f }$. ED of a single unit cell term, $H_f$, and construction of the Hubbard $X$-operators Hubbard $X$-operators $X_f^{pq} = \left| p \right\rangle \left\langle q \right|$ using the complete orthogonal set of eigenstates $\left\{ \left| p \right\rangle \right\}$ of $H_f$.
  \item[\textbf{Step III: Perturbation theory.}] Within the $X$-representation, local interactions are diagonal and all intercell hoppings and long-range Coulomb interaction terms have the bilinear form in the $X$-operators. Various perturbation approaches known for the Hubbard model in the $X$-representation can be used. The most general one includes treatment within the generalized Dyson equation obtained by the diagram technique\cite{r8LDAGTB}.
\end{description}
Below we discuss each step in detail.

\subsection{Step I: LDA}
\label{LDAstep}
LDA provides us a set of Bloch functions $\left| \Psi_{\lambda \vec{k}} \right\rangle$ ($\lambda$ is band index) and band energies $\varepsilon_\lambda(\vec{k})$. For example, LDA band structure calculation for La$_2$CuO$_4$ and Nd$_2$CuO$_4$ was done within the TB-LMTO-ASA (linear muffin-tin orbitals using atomic sphere approximation in the tight-binding) method\cite{r12LDAGTB}. Using the Wannier functions (\textit{W Fs}) formalism\cite{r13LDAGTB} or the NMTO method\cite{r18LDAGTB}, we obtain single electron energies $\varepsilon_\lambda$ and hopping integrals $T_{fg}^{\lambda \lambda'}$ of the TB model\cite{r4LDAGTB,r19LDAGTB}
\begin{eqnarray}
 H &=& \sum\limits_{f,\lambda,\sigma}(\epsilon_ {\lambda}-\mu) n_ {f \lambda \sigma} + \sum\limits_{f \neq g} \sum\limits_ {\lambda,\lambda',\sigma} T_{fg}^{\lambda \lambda'} c_{f \lambda \sigma}^\dag c_{g \lambda' \sigma} \nonumber \\
 &+& \frac {1}{2}\sum\limits_{f,g,\lambda,\lambda'} \sum\limits_{\sigma_1,\sigma_2,\sigma_3,\sigma_4} V_{fg}^{\lambda \lambda'} c_{f \lambda \sigma_1}^\dag c_{f \lambda \sigma_3} c_{g \lambda' \sigma_2}^\dag c_{g \lambda' \sigma_4},
 \label{LDAstep:eq2}
\end{eqnarray}
where $c_{f \lambda \sigma}$ is the annihilation operator in the Wannier representation of the hole at the site $f$ on the orbital $\lambda$ and with the spin $\sigma$,  $n_{f \lambda \sigma }  = c_{f\lambda \sigma }^\dag c_{f\lambda \sigma }$. Note that a number and a symmetry of chosen \textit{W Fs} are determined by the energy window that we are interested in.

The values of Coulomb parameters $V_ {fg}^{\lambda \lambda'}$ are obtained by LDA supercell calculations\cite{r16LDAGTB}. For Cu in La$_2$CuO$_4$, Hubbard parameter $U$ and Hund's exchange $J_H$ are equal to 10 eV and 1 eV, respectively\cite{r17LDAGTB}.

\subsection{Step II: Exact Diagonalization}
\label{EDstep:3.2}
In transition metal (Me) oxides, the unit cell may be chosen as the MeO$_n$ ($n=6,5,4$) cluster and usually there is a common oxygen shared by two adjacent cells. All other ions provide the electroneutrality and contribute to the high energy electronic structure. In the low energy sector, they are inactive. Before ED calculations we solve the problem of nonorthogonality of the oxygen molecular orbitals of adjacent cells. For the $\sigma$-bonding of the $3d$ metal $e_g$ electrons and $a_{1g}$, $b_{1g}$ oxygen orbitals this problem is solved explicitly using the diagonalization in $k$-space\cite{r20LDAGTB}. We have used the same procedure for the $t_{2g}$ orbitals. Such orthogonalization results in the renormalization of the hopping and Coulomb matrix elements in Eq.~(\ref{LDAstep:eq2}). Later we will work with the renormalized parameters. After the orthogonalization, the Hamiltonian~(\ref{LDAstep:eq2}) can be written as a sum of intracell and intercell contributions
\begin{equation}
\label{EDstep:eq3}
H = H_c + H_{cc}, \;\;\; H_c  = \sum\limits_f {H_f}, \;\;\; H_{cc}  = \sum\limits_{f,g} {H_{fg} }
\end{equation}
with orthogonal states in different cells described by $H_f$.

The ED of $H_f$ gives us the set of eigenstates $\left| p \right\rangle  = \left| n_p, i_p \right\rangle$ with the energy $E_p$. Here, $n_p$ is the number of electrons per unit cell and $i_p$ denotes all other quantum numbers like spin, orbital moment, etc. We perform the ED with all possible excited eigenstates, not the Lancoz procedure.

\begin{figure}[t]
\begin{center}
 \includegraphics*[width=0.95\textwidth]{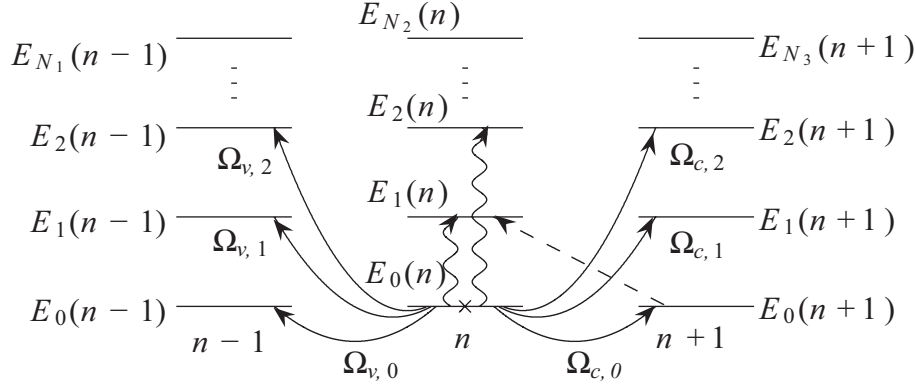}
 \caption{The occupied ground term $E_0(n)$, marked by the cross, and excited terms $E_i(n)$ of the $d^n$ configuration as well as electron removal [addition] $d^{n - 1}$ [$d^{n + 1}$] sectors of the Hilbert space with energies $E_i(n-1)$ [$E_i(n+1)$]. Vertical wavy lines show local Bose-type excitons. Solid lines with arrows show electron removal (index $\nu$) and electron addition (index $c$) Fermi-type excitations. Dashed line shows the virtual Fermi-type excitation from $E_0(n+1)$ to $E_1(n)$.}
\label{fig1}
\end{center}
\end{figure}
How to determine which configurations are relevant? They are found from the local electroneutrality. Let's consider LaMeO$_3$ with Me being a $3d$ element as an example. The ionic valency is La$^{3+}$Me$^{3+}$O$_3^{-2}$, thus the $3d$ cation is Me$^{3+}$ ($d^4$ for Mn$^{3+}$ and $d^6$ for Co$^{3+}$). Due to the covalency, the ground state of the unit cell is given by the hybridization of $d^n p^6  + d^{n + 1} p^5$ configurations (in the spectroscopic notations $d^n + d^{n + 1}\underline{L}$, where $\underline{L}$ means a ligand hole\cite{r21LDAGTB}. Electron addition process results in $d^{n+1}$ subspace of the Hilbert space with mixture of $d^{n + 1} + d^{n + 2}\underline{L}$ configuration. Similarly, electron removal results in $d^{n-1}$ subspace with $d^{n - 1}  + d^n\underline{L}  + d^{n + 1}\underline{L}^2$ mixture. Thus for stoichiometric compound, the three relevant subspaces, $d^{n - 1}$, $d^n$, and $d^{n + 1}$, of the Hilbert space are shown in Fig.~\ref{fig1}. For each subspace, the ED provides a set of multielectron states $\left| n,i \right\rangle$ with energy $E_i (n)$, $i = 0, 1, 2,..., N_n$. Within this set of multielectron terms, the charge-neutral Bose-type excitations with the energy $\omega _i  = E_i (n) - E_0 (n)$ are shown by the vertical wavy lines. Electron addition excitations (local Fermi-type QP) have energies $\Omega _{c,i}  = E_i (n + 1) - E_0 (n)$. Here, index ``c'' means that QPs form the empty conductivity band. Similarly, the valence band is formed by the electron removal Fermi-type QPs with energies $\Omega _{v,i}  = E_0 (n) - E_i (n - 1)$ [a hole creation with energy $E_i (n - 1) - E_0 (n)$ is shown in Fig.~\ref{fig1}]. This multielectron language have been used in the spectroscopy, see for example\cite{r21LDAGTB}. The proper mathematical tool to study both the local QP and their intercell hoppings is given by the Hubbard $X$-operators\cite{r7LDAGTB},
\begin{equation}
\label{EDstep:eq7}
X_f^{pq}  = \left| p \right\rangle \left\langle q \right|.
\end{equation}
with algebra given by the multiplication rule $X_f^{pq} X_f^{rs}  = \delta _{qr} X_f^{ps}$, and by the completeness condition $\sum\limits_p X_f^{pp} = 1$. The last two equations reflects the fact that $X$-operators are the projective operators. The operator $X_f^{pq}$ describes the transition from initial state $\left| q \right\rangle $ to the final state $\left| p \right\rangle $, $X^{pq} \left| q \right\rangle  = \left| p \right\rangle$. The important property of $X$-operators is that any local operator is given by linear combination of $X$-operators. Indeed, $\hat O_f  = \hat 1 \cdot \hat O \cdot \hat 1 = \sum\limits_{p,q} {\left| p \right\rangle } \left\langle p \right|\hat O_f \left| q \right\rangle \left\langle q \right| = \sum\limits_{p,q} {\left\langle p \right|} \hat O_f \left| q \right\rangle X_f^{pq}$. The commutation rule for Hubbard operators follows from the $X$-operators algebra and it is rather awkward. Nevertheless, if $n_p - n_q$ is odd ($\pm 1$, $\pm 3$, etc.) then the $X^{pq}$ is called quasifermionic operator; if $n_p -n_q$ is even ($0$, $\pm 2$, etc.) the $X^{pq}$ is a quasibosonic one\cite{r7LDAGTB}.

We will use a simplified notation
\begin{equation}
 \label{Xm:eq}
 X_f^{pq} \to X_f^{\vec \alpha _m } \to X_f^m,
\end{equation}
where the number $m$ enumerate the excitation and plays the role of the QP band index; each pair $(p,q)$ corresponds to some vector $\vec \alpha(p,q)$ that is called ``root vector'' in the diagram technique\cite{r22LDAGTB}. In this notation, a single-electron (hole) creation operator is given by a linear combination in the $X$-representation, $c_{f\lambda \sigma } = \sum\limits_m \gamma_{\lambda \sigma}\left(m\right) X_{f}^m$, $\gamma_{\lambda \sigma}(p,q) = \left\langle p \right|c_{f\lambda \sigma} \left| q \right\rangle$.

In the $X$-representation, the intracell part of the Hamiltonian is diagonal,
\begin{equation}
\label{EDstep:eq13}
H_c  = \sum\limits_{f,p} \left(E_p - n_p \mu \right) X_f^{pp}.
\end{equation}
The intercell hopping is given by
\begin{equation}
\label{EDstep:eq14}
H_{cc} = \sum\limits_{f \ne g} \sum\limits_{m,m'} t_{fg}^{mm'} X_f^{m\dag} X_g^{m'},
\end{equation}
where the matrix elements are
\begin{equation}
\label{EDstep:eq15}
t_{fg}^{mm'} = \sum\limits_{\sigma,\lambda,\lambda'} T_{fg}^{\lambda \lambda'} \gamma_{\lambda\sigma}^*(m) \gamma_{\lambda'\sigma}(m').
\end{equation}
All intraatomic $d-d$ Coulomb interactions are included in $H_c$ and since $H_c$ is diagonal in $X$-operators, they treated exactly. The dominant part of the $p-d$ and $p-p$ Coulomb interactions is also included in $H_c$ and gives contribution to energies $E_p$, while a small part of it ($\sim 10\%$) provides the intercell Coulomb interaction that is also bilinear in the $X$-operators, $H_{cc}^\mathrm{Coul.} = \sum\limits_{f \ne g} \sum\limits_{p,q,p',q'} V_{fg}^{pq,p'q'} X_f^{pq} X_g^{p'q'}$.

\subsection{Step III: Perturbation Theory}
\label{PTstep:4}
The characteristic local energy scale is given by the effective Hubbard parameter $U_{eff} = E_0 (n + 1) + E_0 (n - 1) - 2E_0 (n)$ that is given by the difference of the initial $d^n+d^n$ and the excited $d^{n-1} + d^{n+1}$ configurations\cite{r21LDAGTB}. The same energy can be obtained as the local gap between the conductivity and valence bands, $U_{eff} = \Omega_{c,0} - \Omega_{v,0}$. Depending on the ratio of the bare Hubbard $U$ and the charge excitation energy $\Delta_{pd} = \varepsilon_p - \varepsilon_d$, the $U_{eff}$ may represent the Mott-Hubbard gap for $U < \Delta_{pd}$ or the charge transfer (CT) gap $E_{CT}$ for $U > \Delta_{pd}$\cite{r23LDAGTB}. The intercell hopping and interaction result in the dispersion and decrease the energy gap, $E_g < U_{eff}$. The intercell hoppings, Eq.~(\ref{EDstep:eq14}), and the non-local Coulomb interactions can be treated by a perturbation theory. We'd like to emphasize that in the $X$-representation the perturbation, Eq.~(\ref{EDstep:eq14}), has exactly the same structure as the hopping Hamiltonian in the conventional Hubbard model. That is why the accumulated experience of the Hubbard model study in the $X$-representation can be used here.

Single-electron Green function for a particle with momenta $\mathbf{k}$, energy $E$, spin $\sigma$, and orbital indices $\lambda$ and $\lambda'$, $G_{\mathbf{k}\sigma}^{\lambda \lambda'}(E) \equiv \left\langle\left\langle c_{\mathbf{k}\lambda\sigma} \left| c_{\mathbf{k}\lambda'\sigma}^\dag \right. \right\rangle\right\rangle_E$, is given by a linear combination of the Hubbard operator's Green functions $D_{\mathbf{k}\sigma}^{mn}(E) = \left\langle \left\langle X_{\mathbf{k}\sigma}^m \left| X_{\mathbf{k}\sigma}^{n\dag} \right. \right\rangle\right\rangle_E$,
\begin{equation}
 \label{EDstep:eq17}
 G_{\mathbf{k}\sigma}^{\lambda\lambda'}(E) = \sum\limits_{mm'} \gamma_{\lambda\sigma}(m) \gamma_{\lambda'\sigma }^*(m') D_{\mathbf{k}\sigma}^{mm'}(E).
\end{equation}

The question is how to determine $X$-operator's Green function. One of the regular ways to treat perturbations is the diagram technique. For $X$-operators with their awkward algebra there is no standard Wick's theorem. Nevertheless, the generalized Wick's theorem has been proved quite some time ago\cite{r24LDAGTB}. The first convenient version of the diagram technique has been formulated in Ref.~\cite{r22LDAGTB}. The general rules of the diagram technique for the $X$-operators are described in details in Ref.~\cite{r8LDAGTB}, where the generalized Dyson equation has been obtained
\begin{equation}
 \label{PTstep:eq28}
 \hat D_{\vec{k}\sigma}(E) = \left\{ \left(E - \Omega_m\right) \delta_{mm'} - \hat P_{\vec{k}\sigma}(E) \hat t(\vec{k}) - \hat \Sigma_{\vec{k}\sigma}(E) \right\}^{-1} \hat P_{\vec{k}\sigma}(E).
\end{equation}
Here $\Omega_m = E_p (n + 1) - E_q (n)$ is the $m$-th QP local energy. Besides the self-energy matrix $\hat \Sigma_{\mathbf{k}\sigma}(E)$, the unconventional term $\hat P_{\mathbf{k}\sigma}(E)$ called ``the strength operator'' appears due to the $X$-operators algebra. Similar term had been known for the spin-operator diagram technique\cite{r25LDAGTB}. This term determines the QP oscillation strength (spectral weight) as well as the renormalized bandwidth. The Hubbard I solution\cite{r7LDAGTB} is obtained by setting $\hat \Sigma_{\mathbf{k}\sigma}(E) = 0$ and $\left(\hat P_{\mathbf{k}\sigma}(E) \right)_{mm'} = F_m \delta_{mm'}$, where $F_m = \left\langle X^{pp} \right\rangle + \left\langle X^{qq} \right\rangle$. We call $F_m$ ``the occupation factor''; it provides a non-zero spectral weight for the QP excitation between at least partially filling eigenstates and gives zero spectral weight for excitations between empty states. The dispersion equation for the QP band structure of the Hubbard fermions in this case is given by
\begin{equation}
\label{PTstep:eq27}
\det \left\| \delta_{mn} \left(E - \Omega_m\right)/F(m) - t^{mn}(\vec{k}) \right\| = 0.
\end{equation}
The dispersion equation, Eq.~(\ref{PTstep:eq27}), is similar to the conventional TB dispersion equation, but instead of a single electron local energy $\varepsilon_\lambda$ we have a local QP energy $\Omega_m$. That is why we call this approach the ``generalized TB method''.

The occupation numbers $\left\langle X^{pp} \right\rangle$ are calculated self-consistently via the chemical potential equation, $n_e = \frac{\left\langle N_e \right\rangle}{N} = \frac{1}{N} \sum\limits_{f,n,i} n\left\langle X_f^{n,i; \; n,i} \right\rangle$. The change of the concentration $n_e$ redistributes the occupation numbers and due to the occupation factors, $F_m$, changes the QP band structure.

In the GTB method, the intracell Green function can be found exactly
\begin{equation}
 \label{EDstep:eq18}
 G_{0 \sigma}^{\lambda\lambda'}(E) = \sum\limits_m \left| \gamma_{\lambda\sigma}(m) \right|^2 \delta_{\lambda\lambda'} D_{0 \sigma}^m(E),
\end{equation}
where $D_{0,\sigma}^m(E) = F_m/ \left(E - \Omega_m  + i\delta\right)$. Comparing this exact local Green function with the Lehmann representation in Eq.~(\ref{QPdef:eq1}), we can say that the electron here is a linear combination of local (Hubbard) fermions with QP energy $\Omega_m$ and a spectral weight $\left| \gamma_{\lambda\sigma}(m) \right|^2 F_m$. It is exactly the same language as in the Lehmann representation. The difference is that it is realized locally and both QP energy and spectral weight are calculated explicitly.

It should be stressed that the LDA+GTB bands are not the single electron conventional bands. There is no any single particle Schr\"odinger equation with the effective potential that gives the LDA+GTB band structure. These QP are excitations between different multielectron terms. The LDA+GTB bands depend on the multielectron term occupation numbers through $\hat P_{\mathbf{k}\sigma}(E)$ and $\hat \Sigma_{\mathbf{k}\sigma}(E)$ that should be calculated via the chemical potential equation. There is no rigid band behavior from the very beginning; the band structure depends on doping, temperature, pressure, and external fields.

\section{Cuprates}
\label{CalcEx:5}
In this section we present the results for cuprates with one CuO$_2$ layer in the unit cell: hole doped cuprate La$_{2 - x}$Sr$_x$CuO$_4$. Yet there is no self-consistent treatment of both electronic and magnetic structures; our LDA+GTB calculation have been carried out for a prescribed magnetic order. Thus, we consider separately the antiferromagnetic (AFM) phase with the long-range AFM order and the spin-liquid phase with the short-range AFM correlations.

\subsection{Band Structure of the Undoped La$_2$CuO$_4$}
\label{CalcEx:5.1}
The ED of the multiband $p-d$ Hamiltonian~(\ref{LDAstep:eq2}) for the CuO$_6$ cluster which includes the apical oxygens, results in the following local eigenstates (in the hole representation): 1) $n_h = 0$, the vacuum state $\left| 0 \right\rangle$ formed by $d^{10} p^6$ orbital configuration, 2) $n_h = 1$, the spin doublets $\left| \sigma,\lambda \right\rangle$ with different orbital symmetries. The lowest one is $b_{1g}$, $\left| \sigma \right\rangle$, and the first excited is $a_{1g}$ molecular orbital. 3) $n_h = 2$. A set of two-hole singlets and triplets, spread in the energy region of about $U_d \sim 10$ eV. The lowest one is the $^1 A_1$ singlet $\left| S \right\rangle$ that includes the Zhang-Rice singlet among other two-hole singlets. The first excited triplet $\left| TM \right\rangle$ ($M=+1, 0, -1$) has the $^3 B_{1g}$ symmetry. The total number of eigenstates is about 100.

\begin{figure}[t]
\begin{center}
 \includegraphics*[width=0.9\textwidth]{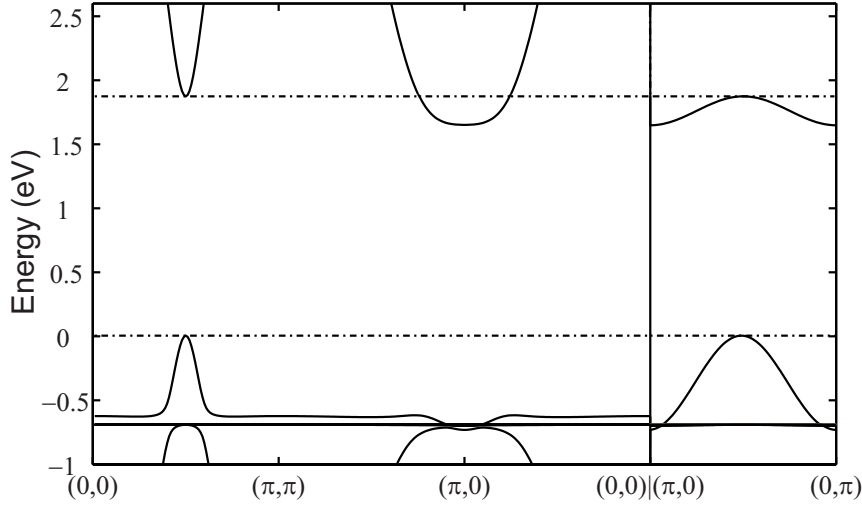}
 \caption{The LDA+GTB band structure of the AFM La$_2$CuO$_4$ along the principal cuts of the Brillouin zone \protect\cite{r4LDAGTB}.
\label{fig3}}
\end{center}
\end{figure}
The next practical step is the calculation of the matrix elements,
$\left\langle 0 \right| c_{f \lambda \sigma} \left| {1\sigma' \lambda} \right\rangle$, $\left\langle {1 \sigma' \lambda} \right|c_{f \lambda \sigma} \left| {2,i} \right\rangle$, and construction of $X$-operators for all single electron orbitals. Here, $\lambda$ stands for Cu-$d_{x^2 - y^2}$, Cu-$d_{z^2}$, O-$b$, O-$a$, or O-$p_z$ orbital. For the AFM-ordered La$_2$CuO$_4$, we use the two-sublattice ($A$ and $B$) version of the Hubbard I solution with two occupation factors, $F_{m,A}$ and $F_{m,B}$. Due to the effective molecular field, the local $b_{1g}$ spin doublet is splitted so that at $T = 0$: $\left\langle X_A^{\uparrow\uparrow} \right\rangle = \left\langle X_B^{\downarrow\downarrow} \right\rangle = 1$, $\left\langle X_A^{\downarrow\downarrow} \right\rangle = \left\langle X_B^{\uparrow\uparrow} \right\rangle = 0$. The GTB band structure and the DOS in the wide energy region with all excited two-hole states $\left| {2,i} \right\rangle$ have been calculated in Ref.~\cite{r28LDAGTB}. The empty conductivity band is formed by only one Hubbard fermion, $X_f^{0,\sigma}$. It is separated by the CT gap $U_{eff} = E_{CT} \approx 2$ eV from the filled valence band. The valence band is formed by a large number of Hubbard fermions $X_f^{\sigma,2i}$ and consists of a set of narrow bands with the total bandwidth about $6$ eV. If we are interested in a smaller energy window around the $E_{CT}$ (for example, to study ARPES), it is possible to simplify the calculation by neglecting the high-energy states from both $\left| 2,i \right\rangle$ and $\left| 1 \sigma' \lambda \right\rangle$ sets. Then the minimal realistic basis is $\left\{ \left| 0 \right\rangle, \left| \sigma \right\rangle, \left| S \right\rangle, \left| TM \right\rangle \right\}$.

The $X$-representation for the fermionic operators in this basis is
$c_{f d_{x^2-y^2} \sigma} = u X_f^{0 \sigma}  + 2 \sigma \gamma_x X_f^{\bar \sigma S}$, $c_{f p_b \sigma} = v X_f^{0 \sigma}  + 2 \sigma \gamma_b X_f^{\bar \sigma S}$, $c_{f p_a \sigma} = \gamma_a (\sigma \sqrt{2} X_f^{\bar\sigma T0} - X_f^{\sigma T2\sigma})$, $c_{f d_{z^2} \sigma} = \gamma_z (\sigma \sqrt{2} X_f^{\bar\sigma T0} - X_f^{\sigma T2\sigma})$, $c_{f p_z \sigma} = \gamma_p (\sigma \sqrt{2} X_f^{\bar\sigma T0} - X_f^{\sigma T2\sigma})$. Here, $\bar{\sigma} \equiv -\sigma$ and $T2\sigma$ stands for $T(+1)$ or $T(-1)$ depending on the value of the spin label $\sigma=\pm 1/2$. The explicit form of the TB Hamiltonian~(\ref{LDAstep:eq2}) in this basis looks like the two-band singlet-triplet Hubbard model:
\begin{eqnarray}
H_{pd} &=& \sum\limits_{f} \left[ \varepsilon_{1} \sum_{\sigma} X_{f}^{\sigma \sigma} + \varepsilon_{2S}X_{f}^{S S} + \varepsilon_{2T} \sum_{M} X_{f}^{TM TM} \right] \nonumber \\
&+& \sum\limits_{f \neq g, \sigma} \left[
t_{fg}^{0 0} X_{f}^{\sigma 0} X_{g}^{0 \sigma} + t_{fg}^{S S} X_{f}^{S \bar\sigma} X_{g}^{\bar\sigma S} + 2 \sigma t_{fg}^{0 S} \left( X_{f}^{\sigma 0} X_{g}^{\bar\sigma S} + h.c. \right) \right. \nonumber \\
&+& t_{fg}^{S T} \left\{ \left( \sigma \sqrt{2} X_{f}^{T0 \bar\sigma} - X_{f}^{T2\sigma \sigma} \right) \left( v X_{g}^{0 \sigma} + 2\sigma \gamma_{b} X_{g}^{\bar\sigma S} \right) + h.c. \right\} \nonumber \\
&+& \left. t_{fg}^{T T} \left( \sigma \sqrt{2} X_{f}^{T0 \bar\sigma} - X_{f}^{T2\sigma \sigma} \right) \left( \sigma \sqrt{2} X_{g}^{\bar\sigma T0} - X_{g}^{\sigma T2\sigma} \right) \right].
\label{CalcEx:eq32}
\end{eqnarray}
The hopping parameters, Eq.~(\ref{EDstep:eq15}), of the effective Hubbard model are expressed through the microscopic \textit{ab initio} parameters
$t_{pd}$ and $t_{pp}$, $t_{fg}^{00} = -2t_{pd}\mu_{fg}2uv - 2t_{pp}\nu_{fg}v^2$, $t_{fg}^{SS} = -2t_{pd}\mu_{fg}2\gamma_x\gamma_b - 2t_{pp}\nu_{fg}\gamma_b^2$, $t_{fg}^{0S} = -2t_{pd}\mu_{fg}(v\gamma_x+u\gamma_b) - 2t_{pp}\nu_{fg}v\gamma_b$, $t_{fg}^{TT} = \frac{2}{\sqrt{3}}t_{pd}\lambda_{fg}2\gamma_a\gamma_z + 2t_{pp}\nu_{fg}\gamma_a^2 - 2t'_{pp}\lambda_{fg}2\gamma_p\gamma_a$, $t_{fg}^{ST} = \frac{2}{\sqrt{3}}t_{pd}\xi_{fg}\gamma_z + 2t_{pp}\chi_{fg}\gamma_a - 2t'_{pp}\xi_{fg}\gamma_p$. Here $\mu_{fg}$, $\nu_{fg}$, $\lambda_{fg}$, $\xi_{fg}$, and $\chi_{fg}$ are the coefficients of the oxygen group orbitals construction, and $u$, $v$, $\gamma_x$, $\gamma_b$, $\gamma_a$, $\gamma_p$, and $\gamma_z$ are the matrix elements $\gamma_{\lambda\sigma}(p,q)$ (see Ref.~\cite{r2LDAGTB} for details).

The QP band structure of La$_2$CuO$_4$ is shown in Fig.~\ref{fig3} for the $\Gamma(0,0) - M(\pi,\pi) - X(\pi,0) - \Gamma(0,0)$ and $X(\pi,0) - Y(0,\pi)$ cuts of the square lattice Brillouin zone. Zero at the energy scale is not the Fermi level but rather fixed by the condition $\varepsilon_{d_{x^2 - y^2}} = 0$.

The top of the valence band is at the $\bar M = (\pi/2,\pi/2)$ point, while the bottom of the conductivity band is at the $X$ point. The dispersion of the valence band determined by the hybridization of the two bands which formed by either $X_f^{\bar\sigma S}$ or $X_f^{\bar\sigma TM}$ Hubbard fermions. The hybridization between them is provided by the $t^{ST}$ hopping matrix elements in Eq.~(\ref{CalcEx:eq32}). These are fermionic bands, but frequently in the literature terms ``singlet band'' and ``triplet band'' are used. These terms reflect the final two-hole states involved in the QP excitations. The dominant spectral weight in the singlet band stems from the oxygen $b_{1g}$ states, while for the bottom of the conductivity band it is from the $d_{x^2 - y^2}$ states of Cu.

\subsection{Cascade of Lifshitz Quantum Phase Transitions in La$_{2 - x}$Sr$_x$CuO$_4$}
\label{CalcEx:5.3}
The low-energy Hamiltonian for La$_{2 - x}$Sr$_x$CuO$_4$ (LSCO) is the $t - t' - t'' - J^*$ model obtained via exclusion of the interband (through the charge-transfer gap) excitations. Here $J^*$ means that besides the Heisenberg exchange term a three-site correlated hopping $H_3$ is also included, $H_{t - J^*} = H_{tJ} + H_3$, where
\begin{eqnarray}
\label{eq4}
 H_{tJ} &=& \sum\limits_{f, \sigma} (\varepsilon - \mu) X_f^{\sigma \sigma}
 + \sum\limits_f 2(\varepsilon - \mu) X_f^{SS} \nonumber\\
 &+& \sum\limits_{f \ne g, \sigma} \left[ t_{fg} X_f^{S \bar\sigma} X_g^{\bar\sigma S}
 + \frac{J_{fg}}{4} \left( X_f^{\sigma \bar\sigma} X_g^{\bar\sigma \sigma} - X_f^{\sigma \sigma} X_g^{\bar\sigma \bar\sigma} \right) \right], \nonumber\\
 H_3 &=& \sum\limits_{f \ne m \ne g, \sigma} \frac{\tilde t_{fm} \tilde t_{mg}}{U_{eff}}  \left( X_f^{\sigma S} X_m^{\bar\sigma \sigma} X_g^{S \bar\sigma} - X_f^{\sigma S} X_m^{\bar\sigma \bar\sigma} X_g^{S \sigma} \right). \nonumber
\end{eqnarray}
Here $t_{fg} = t_{fg}^{SS}$ is the hopping in the UHB, $J_{fg} = \tilde{t}_{fg}^2/U_{eff}$ is the exchange interaction due to the interband (UHB~$\leftrightarrow$~LHB) hopping $\tilde t_{fg} = t_{fg}^{0S}$ through the charge-transfer gap $U_{eff} = E_{CT}$, hole creation operator is now $\tilde a_{f\sigma}^\dag = 2\sigma X_f^{S \bar\sigma}$ and its algebra is different from the bare fermion's one. The spin operators are also easily expressed via $X$-operators, $S_f^+= X_f^{\sigma
\bar\sigma}$, $S_f^z = \left( X_f^{\sigma \sigma} - X_f^{\bar\sigma \bar\sigma} \right)/2$.

Here we dropped out the triplet state $\left| TM \right>$ because the triplet itself and the singlet-triplet excitations do not contribute to the near-Fermi level physics.

The \textit{ab initio} derived parameters for LSCO are (in eV)
$t = 0.93$, $t' = -0.12$, $t'' = 0.15$, $\tilde t = 0.77$, $\tilde t' = -0.08$, $\tilde t'' = 0.12$, $J = 0.29$, $J' = 0.003$, $J'' = 0.007$.

Our approach is essentially a perturbation theory with the small parameter
$t/U$ contrary to the usual Fermi liquid perturbation expansion in terms of $U$ which is large in cuprates. We use a method of irreducible Green functions
which is similar to the Mori-type projection technique, with the zero-order
Green function given by the well-known Hubbard I approximation. Beyond it there are spin fluctuations. To provide a description of them, the self-energy was
calculated in the non-crossing approximation by neglecting vertex
renormalization that is equivalent to the self-consistent Born approximation
(SCBA)\cite{r23}. Resulting electron self-energy contains the space-time
dependent spin correlation function $C(\mathbf{q}, \omega)$ and results in the
finite quasiparticle lifetime, $\mathrm{Im}\Sigma(\mathbf{k},\omega) \ne 0$.
Note that at low temperatures $T \le 10$K the spin dynamics is much slower than the electron one. A typical spin fluctuation time, $10^{-9}$ sec, is much
larger than the electronic time $10^{-13}$ sec\cite{r26}; that is why we can
safely neglect the time dependence of the spin correlation function,
$C(\mathbf{q},\omega) \to C_\mathbf{q}$. The self-energy becomes static,
$\Sigma(\mathbf{k}, \omega) \to \Sigma(\mathbf{k})$, and we have
$\mathrm{Im}\Sigma = 0$. Note that $\Sigma(\mathbf{k}, \omega)$ here is the
object completely different from the one in the Fermi liquid approach because
the here it is build by the diagrams for the $X$-operators, not the
standard Fermionic annihilation-creation operators $a_{f\sigma}$. In the usual
Fermi liquid expansion dynamical self-energy definitely plays a crucial role in the lightly doped cuprates. Here, our theory starts from a different limit
where the lowest order approximation is represented by the Hubbard I solution.
The corrections to the strongly-correlated mean-field approach are small
because the starting point is already a reasonable approximation for the
Mott-Hubbard insulator. That is proved by the small effect of the frequency
dependence of the self-energy in Refs.~\cite{r23,r22}. Moreover the doping-dependence of the FS is determined by $\mathrm{Re}\Sigma$, and it is
qualitatively similar in our approach\cite{r25} and in the approach which
properly takes $\mathrm{Im}\Sigma$ into account\cite{r23,r22}.

The vertex corrections to the self-energy are small far from the spin-density
wave or the charge-density wave instabilities, that is true for the moderate
doping. Our approximation for the self-energy is done in the framework of the
mode-coupling approximation which has been proved to be quite reliable even for systems with strong interaction\cite{r40,r41}. As shown in the spin-polaron
treatment of the $t-J$ model, the vertex corrections to the non-crossing
approximation are small and give only numerical renormalization of the model
parameters\cite{r42}.

Green function $\left\langle \left\langle \left. X_\mathbf{k}^{\bar\sigma S}
\right| X_\mathbf{k}^{S \bar\sigma} \right\rangle \right\rangle_\omega$ for a
hole moving on the background of short-range AFM order is
\begin{eqnarray}
\label{eq5}
 G(\mathbf{k},\omega) = \frac{(1+x)/2}{\omega - \varepsilon + \mu - \frac{1 + x}{2} t_\mathbf{k} - \frac{1 - x^2}{4} \frac{\tilde{t}_\mathbf{k}^2}{U_{eff}} + \Sigma(\mathbf{k})},
\end{eqnarray}
where
\begin{eqnarray}
\label{eq6}
 \Sigma(\mathbf{k}) &=& -\frac{2}{1 + x} \frac{1}{N} \sum\limits_\mathbf{q} \left\{ \left[ t_{\mathbf{k} - \mathbf{q}} - \frac{1 - x}{2} J_\mathbf{q}  + \frac{1 - x}{2} \frac{\tilde{t}_{\mathbf{k} - \mathbf{q}}^2}{U_{eff}} \right.\right. \nonumber\\
 &-& \left.\left. \frac{1 + x}{2} \frac{2 \tilde{t}_\mathbf{k} \tilde{t}_{\mathbf{k} - \mathbf{q}}}{U_{eff}} \right] \left( \frac{3}{2}C_\mathbf{q} + K_{\mathbf{k} - \mathbf{q}} \right) - \frac{1 + x}{2} \frac{\tilde{t}_\mathbf{q}^2}{U_{eff}} K_\mathbf{q} \right\}.
\nonumber
\end{eqnarray}
Here, $t_\mathbf{k}$ and $\tilde{t}_\mathbf{k}$ are the Fourier transforms of
hoppings $t_{fg}$ and $\tilde t_{fg}$, respectively. The self-energy is determined by static spin correlation function $C_{0n} = \left\langle {S_0^+ S_n^-} \right\rangle$ and kinetic correlation function $K_{0n} = \sum\limits_\sigma \left\langle \tilde a_{0\sigma }^\dag \tilde a_{n\sigma} \right\rangle$ between sites 0 and $n$. These correlation functions and their Fourier transforms $C_\mathbf{q}$ and $K_\mathbf{q}$ represent the AFM short-range order and the valence-bond order, respectively. In contrast to
approach of Ref.~\cite{r23}, we calculate these correlation functions
self-consistently up to $n = 9$  (ninth coordination sphere) together with the
chemical potential $\mu$. To get the spin correlation function we also obtain
the spin Green function $\left\langle \left\langle \left. X_\mathbf{q}^{\sigma
\bar\sigma} \right| X_\mathbf{q}^{\bar\sigma \sigma} \right\rangle
\right\rangle_\omega$ in a spherically-symmetric spin liquid
state~\cite{rr30,r24} with $\left\langle S^z \right\rangle = 0$ and the equal
correlation functions for each spin component, $\left\langle S_0^+ S_n^-
\right\rangle = 2 \left\langle S_0^z S_n^z \right\rangle = C_{0n}$. Both
$C_{0n}$ and $K_{0n}$ are essentially doping-dependent and $C_{0n}$ decrease
with the doping~\cite{r25}. While the nearest neighbor function $C_{01}$ is finite for all studied $x$ up to $x = 0.4$ with a kink at $x = p^* = 0.24$, more distant spin correlations fall down to zero for $x > p^*$.

\begin{figure}
\begin{center}
 \includegraphics[angle=0,width=0.75\columnwidth,clip=true]{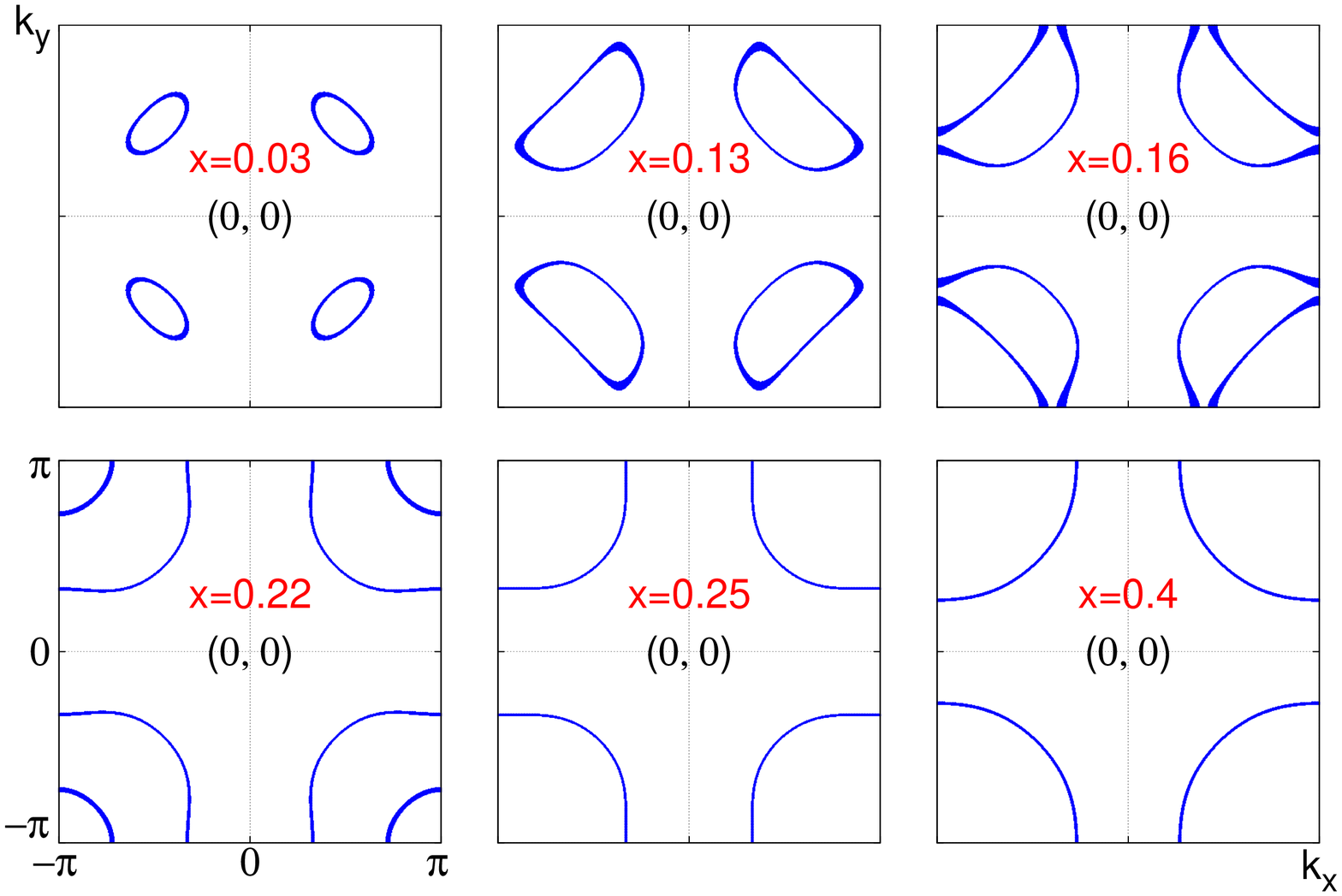}
 \caption{(Color online) Mean-field Fermi surface transitions with doping $x$ as calculated from poles of Eq.~(\ref{eq5}). There are two topological changes: first one between $x=0.13$ and $0.16$, and second one between $0.22$ and $0.25$; see Ref.~\protect\cite{r25} for detailed discussion.}
\label{figFS}
\end{center}
\end{figure}
The calculated FS twice changes its topology with doping\cite{r25}, see
Fig.~\ref{figFS}. Small hole pockets around $(\pm \pi/2, \pm \pi/2)$ points are present at small doping; then they increase in size and touch each other in the non-symmetric points $\mathbf{k} = \pm \pi (1, \pm 0.4)$ at $x_{c1} = p_{opt} = 0.151$. Above $p_{opt}$, there are two FSs around $(\pi, \pi)$ with outer being a hole-like and inner being an electron-like. The electron FS collapsed at
$x_{c2} = p^* = 0.246$, and at $x > p^*$ we have only one large hole surface
around $(\pi, \pi)$. Similar conclusion on the coexistence of hole and electron FS at some intermediate doping have been also drawn recently\cite{r27,r28},
and earlier for the spin-density wave sate of the Hubbard
model~\cite{rSachdev}.

From this consideration it follows that the FS topological transitions in cuprates are induced by doping and they are due to the non-rigid band behavior of the quasiparticles in systems with SEC. According to the general Lifshitz analysis\cite{r8} for the three dimensional (3D) system, a change of topology at the energy $\varepsilon = \varepsilon_c$ either by appearance of a new segment (like we found at $p^*$) or by change of its connectivity (like at $p_{opt}$) would result in the additional DOS, $\delta N(\varepsilon) \sim \left( \varepsilon - \varepsilon_c \right)^{1/2}$, and the change in the thermodynamic potential, $\delta\Omega \sim \left(\varepsilon_F - \varepsilon_c \right)^{5/2}$ (the QPT of the 2.5-order), where $\varepsilon_F$ is the Fermi energy. However, due to the strong anisotropy of electronic and magnetic properties, cuprates are quasi-2D and not isotropic 3D systems. The electron hopping perpendicular to the CuO$_2$ layers in a single-layer cuprates like LSCO is negligibly small.

\begin{figure}
\begin{center}
 \includegraphics[angle=0,width=0.7\columnwidth]{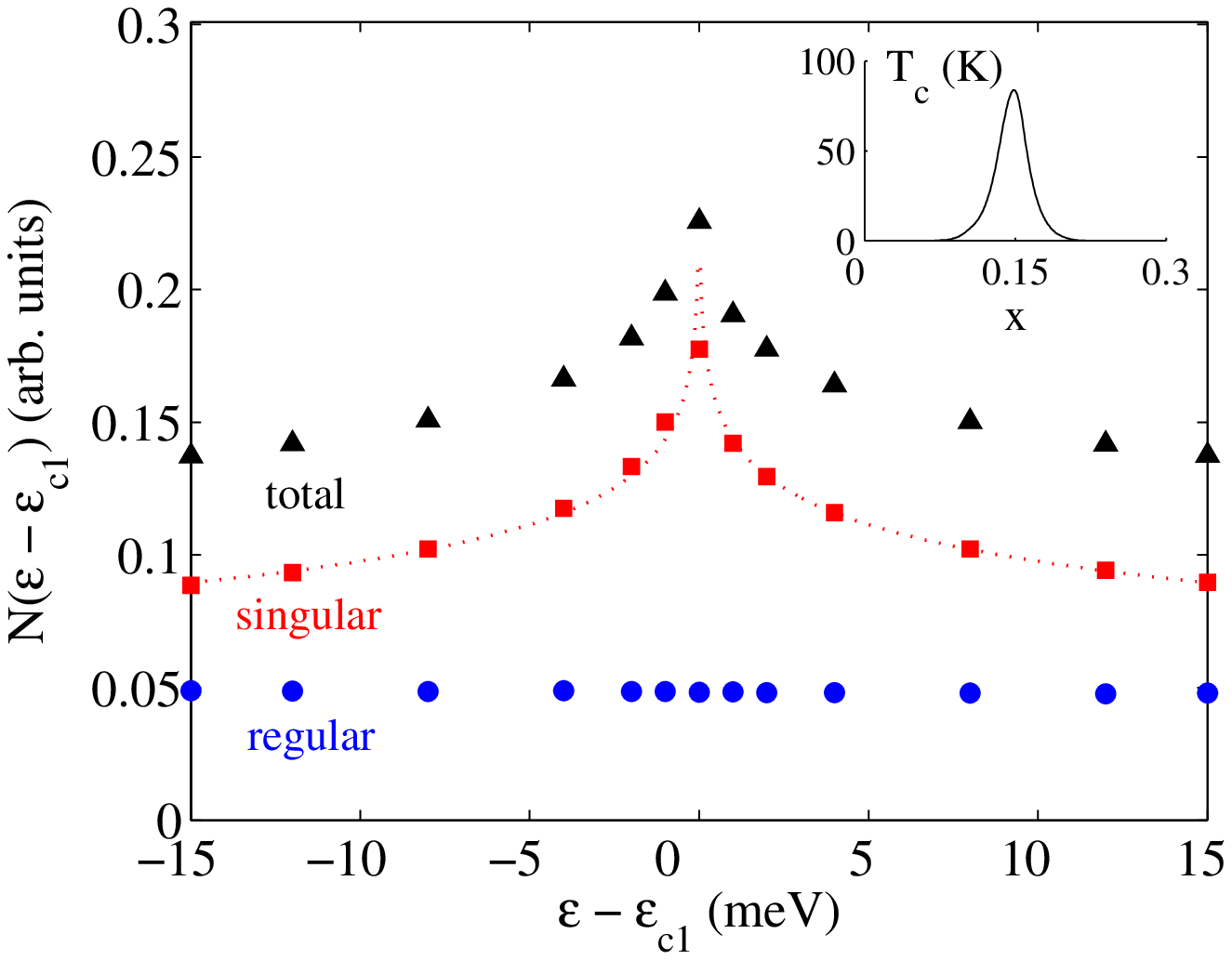}
 \includegraphics[angle=0,width=0.7\columnwidth]{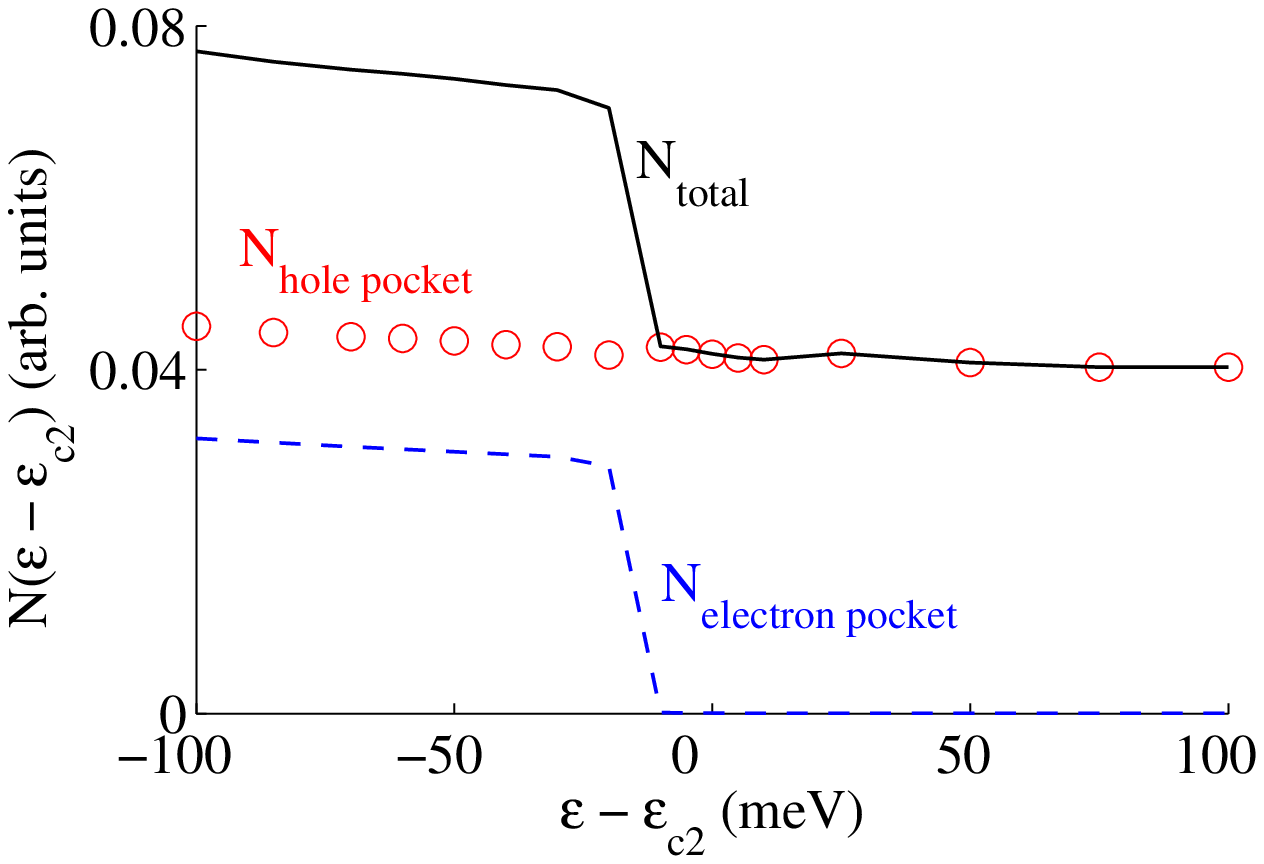}
 \caption{Regular, singular, and total density of states $N\left(\varepsilon-\varepsilon_{ci}\right)$ near the optimal doping $\varepsilon_{c1}=\varepsilon_F(p_{opt})$ [top] and near the pseudogap critical point $\varepsilon_{c2} = \varepsilon_F(p^*)$ [bottom], as calculated from the Green function (\ref{eq5}). Dotted curve shows the logarithmic fitting. Below $p^* = 0.24$ ($\varepsilon < \varepsilon_{c2}$) a singular step-like contribution to the total DOS appears due to the appearance of the electron pocket. In the inset, the doping dependence of the superconducting critical temperature $T_c(x)$ is shown; the optimal doping is $0.151$. Note that the energy $\varepsilon - \varepsilon_{c1}$ is the energy of holes.}
\label{figdos}
\end{center}
\end{figure}
The change of the FS topology at $x_{c1} = p_{opt}$ results in the logarithmic
divergence of DOS, while the emergence of the new
electron-like pocket below $x_{c2} = p^*$ results in a step in DOS
(Fig.~\ref{figdos}). The total DOS is a sum of the singular and regular
contributions. We would like to stress that both logarithmic and step DOS singularities are in perfect agreement with the general properties of the van Hole singularities for the 2D electrons\cite{rr35}. Contrary to the 3D systems, the thermodynamical potential for the 2D electrons has a singular contribution $\delta\Omega \sim \left( \varepsilon_F - \varepsilon_c \right)^2$ for the step singularity and $\delta\Omega \sim \left( \varepsilon_F - \varepsilon_c \right)^2 \ln\left| \varepsilon_F - \varepsilon_c \right|$ for the logarithmic singularity\cite{r29}. Thus QPT at $x_{c2} = p^*$ is of the second order, while at $x_{c1} = p_{opt}$ the singularity is stronger. It is immediately follows that the Sommerfeld parameter $\gamma$ in the electronic heat capacity $\gamma = C_e/T$ has also a singular step contribution at $x \le p^*$, and $\delta\gamma \propto \ln\left( \varepsilon_F - \varepsilon_c \right) \propto \ln\left| x - x_{opt} \right|$ near $x_{c1} = p_{opt}$. Similar divergence in the specific heat was found within the dynamical cluster approximation for the Hubbard model\cite{Mikelsons2009}.

The coincidence of $x_{c1}$ with $p_{opt}$ and $x_{c2}$ with
$p^*$ is not occasional. We compare the superconducting critical
temperature dependence $T_c(x)$ in the same model\cite{r30} as a function of doping and observe that $T_c(x)$ has a \textit{maximum} at $x_{opt}$ (see inset in Fig.~\ref{figdos}), which indeed equals to $x_{c1}$. It is not a coincidence since like in the BCS theory the maximum in $T_c(x)$ is determined by the maximum DOS, and at $x_{c1}$ we have a logarithmic singularity. Kinetic energy, $E_{kin} = \sum\limits_n t_{0n} K_{0n}$, reveals a remarkable kink at $x_{c2} = p^*$ due to the change in DOS. Above $p^*$, $E_{kin}(p)/E_{kin}(p^*) \sim 1 + p$ that is expected for a conventional 2D metal with the hole concentration $n_h = 1 + p$ and $E_{kin} \sim \varepsilon_F \sim n_h$. The extrapolation of this law below $p^*$ reveals that actual $E_{kin}$ is much smaller. We associate this depletion with the \textit{pseudogap formation} and it fits very well with the Loram-Cooper model\cite{rr1,r1} -- a simple free electron gas with a triangular pseudogap DOS,
\begin{equation}
\label{eq2}
 N(\varepsilon) = \left\{ \begin{array}{l}
              g, \left| \varepsilon - \varepsilon_F \right| > E_g \\
              g \frac{\left| \varepsilon - \varepsilon_F \right|}{E_g},
               \left| \varepsilon - \varepsilon_F \right| < E_g.
                         \end{array} \right.
\end{equation}
Here $E_g = J (p^* - p)/p^*$ is a doping dependent pseudogap and $J$ is the
nearest-neighbor exchange parameter. The observed depletion of the kinetic energy together with the jump in DOS relates the QPT at $x_{c2}$ to the pseudogap and the coincidence of $x_{c2}$ and $p^*$ is not occasional.

The energy dependence of the electron self-energy is crucial and determines the Mott-Hubbard transition in the Hubbard model as was convincingly demonstrated
by the dynamical mean-field theory (DMFT)\cite{r43}. Cluster generalization of DMFT\cite{Hettler_1998,Kotliar_2001,Potthoff_2003,Maier_2005} is necessary to study electron correlations in a two-dimensional CuO$_2$ layer where the nearest neighbor spin correlations require the momentum dependent self-energy. The cellular DMFT (CDMFT) method provides $k$-dependent self-energy and results in the phase
diagrams that have features similar to the ones experimentally observed in
cuprates\cite{r48,r49,r50,r51,r52}. Recently, the exact diagonalization
version of CDMFT (CDMFT+ED) was used to study the electronic structure of the doped Mott-Hubbard insulator\cite{r53,r54}. The sequence of the FS transformations with doping in Refs.~\cite{r53,r54} is very similar
to ours and this is an additional prove of the our approach's validity at least at low temperatures and away from the Fermi-liquid regime.

\section{Band Structure of Manganite La$_{1 - x}$Sr$_x$MnO$_3$}
\label{CalcEx:6}
The most unusual solutions of Eq.~(\ref{PTstep:eq27}) are those with the zero spectral weight ($F_m  = 0$). We know the QP energy and this QP has zero number of states. We call such QP the ``virtual QP''. To obtain the non-zero spectral weight $F_m \ne 0$ the non-zero occupation of the initial or final states for the excitation $\alpha_m = (p,q)$ is required. It may be achieved by doping, pressure, finite temperature, and external field. The virtual states have been found in GTB calculations for La$_2$CuO$_4$\cite{r2LDAGTB} and for FeBO$_3$\cite{r26LDAGTB}. Recently they have been experimentally observed in FeBO$_3$ by the IR (infra-red) spectroscopy\cite{r27LDAGTB}.

\begin{figure}[t]
\begin{center}
 \includegraphics*[width=1.0\textwidth]{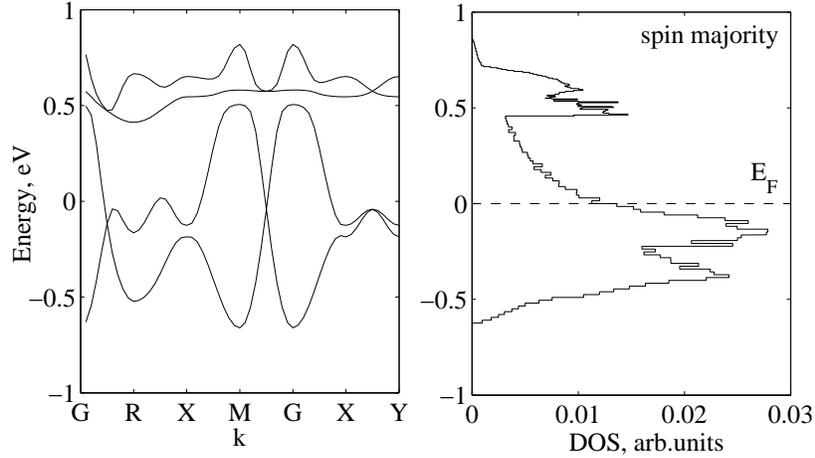}
 \caption{LDA+GTB dispersion of the in-gap and valence bands (on the left) and spin majority density of states (on the right) for La$_{1 - x}$Sr$_x$MnO$_3$ in the ferromagnetic phase ($x = 0.3$). The spin minority DOS in the same energy region is zero. Here, $G=(0,0,0)$, $R=(\pi,\pi,\pi)$, $X=(\pi,0,0)$, $M=(\pi,\pi,0)$, and $Y=(0,\pi,0)$ are the main symmetry points of the Brillouin zone.}
 \label{fig6}
\end{center}
\end{figure}

Band structure calculations for manganites is more complicated than for cuprates for two reasons: a) the orbital ordering double the unit cell, b) the high spin values for $d^4$ ($S=2$) and for $d^{4 \pm 1}$ ($S=5/2$ for $d^5$ and $S=3/2$ for $d^3$) increase the number of states in the Hilbert space. The construction of the $X$-operators representation for LaMnO$_3$ within the high spin $d^3$, $d^4$, and $d^5$ configurations has been done in Ref.~\cite{r5LDAGTB}. For the undoped LaMnO$_3$, the LDA+GTB calculations result in the QP band structure with the large CT gap $E_{CT} \approx 2$ eV. Above the top of the occupied valence band there is a virtual state with the activation energy $\Delta\varepsilon \approx 0.4$ eV. Doping or non-stoichiometry transforms this state into the in-gap narrow band. 
The LDA+GTB calculations for hole-doped La$_{1 - x}$Sr$_x$MnO$_3$ for $x = 0.2 \div 0.3$ resulted in the half-metallic ground state with the 100\% spin polarization at the Fermi level in the ferromagnetic phase, that is illustrated in Fig.~\ref{fig6}.

Transition from the ferromagnetic to the paramagnetic phase results in
the band narrowing $\propto \frac{1}{2}\cos(\theta)$ at $x = 0$ that is two times stronger than the double exchange model provides\cite{r40LDAGTB}.

\section{Finite Temperature Effect on the Electronic Structure of LaCoO$_3$}
\label{lacoo3:7}
The perovskite-oxide LaCoO$_{3}$ has been studied intensely for many years
due its unique magnetic properties and related insulator-metal transition
(IMT)\cite{orlov1,orlov2}. A gradual appearance of the paramagnetism above 50K from the diamagnetic ground state in called the spin-state transition. Goodenough was the first who suggested that instead of the Hund's rule dictated high spin (HS) $S=2$ the strong crystalline electric field results in the low spin (LS) S=0 state for $d^{6}$ configuration of the Co$^{3+}$ ion, and the energy difference is very small with the spin gap $\Delta_{S} = E(\mathrm{HS}) - E(\mathrm{LS}) \sim 100$K. The thermal population of the HS state provides the sharp increase of the magnetic susceptibility $\chi$ with a maximum around 100K. The nature of the excited spin state of Co$^{3+}$ above the singlet
$^{1}A_{1g}$ has been under debate (see for a recent review\cite{orlov3}). Besides the original $^{5}T_{2g}$ HS state with the $t^{4}_{2g}e^{2}_{g}$
configurations\cite{orlov2} there were many indications on the intermediate spin (IS) $S = 1$, $^{3}T_{1g}$, state. The two stage model has been proposed with the LS-IS transition at 100K and IS-HS transition at 550-600K\cite{orlov4,orlov5}. Recent electronic spin resonance (ESR)\cite{orlov6}, X-ray absorption spectroscopy (XAS) and X-ray magnetic circular dichroism (XMCD)\cite{orlov7} experiments prove that the lowest excited state is really the HS. Nevertheless the $^{5}T_{2g}$ term is splitted by the spin-orbital interaction in the low energy triplet with effective moment $J = 1$, and higher energy sublevels with $J = 2$ and $J = 3$\cite{orlov8}. The large difference between the spin excitation gap $\Delta_{S}$ and the charge gap given by the activation energy for electrical conductivity $E_{a} \approx 0.1$eV at low $T$ indicates that LaCoO$_{3}$ is not a simple band insulator\cite{orlov9}. The second shallow maximum in $\chi$ near $500 \div 600$ is often related to the insulator-metal transition. Surprisingly for the IMT electrical conductivity $\sigma$ does not seem to show any noticeable anomaly at this temperature\cite{orlov9}. Moreover the discrepancy between the large charge gap $2E_{a} \approx 2300$K and the $T_\mathrm{IMT} \approx 600$K implies that the IMT cannot be simply argued in terms of a narrow-gap semiconductor\cite{orlov10}. We solved this problem by calculating the electronic band structure in the regime of strong electron correlations. We consider electron as the linear combination of quasiparticles (QP) given by excitations between the different multielectron configurations obtained by exact diagonalization of the CoO$_{6}$ cluster. With the Hubbard operators constructed within the exact cluster eigenstates we can calculate the QP band structure for the infinite lattice. The QP spectral weight is determined by the occupation numbers of the local multielectron configurations. We find that the thermal population of different sublevels of the $^{5}T_{2g}$ HS term splitted by the spin-orbital interaction results both in the spinstate transition and also in some new QP excitations. Of particular importance is the hole creation QP from the initial $d^{6}$ HS into the $d^{5}$ HS term, this QP appears to
form the in-gap state inside the large charge-transfer gap $E_{g} \approx 1.5$eV. The intercluster hopping transforms this local QP into the in-gap
band that lies just under the bottom of empty conductivity band and provides
the insulating gap $2E_{a} \approx 0.2$eV at $T = 100$K. It bandwidth
increases with $T$, and overlapping with the conductivity band at $T = T_\mathrm{IMT} = 587$K results in the IMT. Hence our approach allows to treat both the low $T$ spin-state transition and the high T IMT on the same footing.

\begin{figure}[t]
\begin{center}
 \includegraphics*[width=1.0\textwidth]{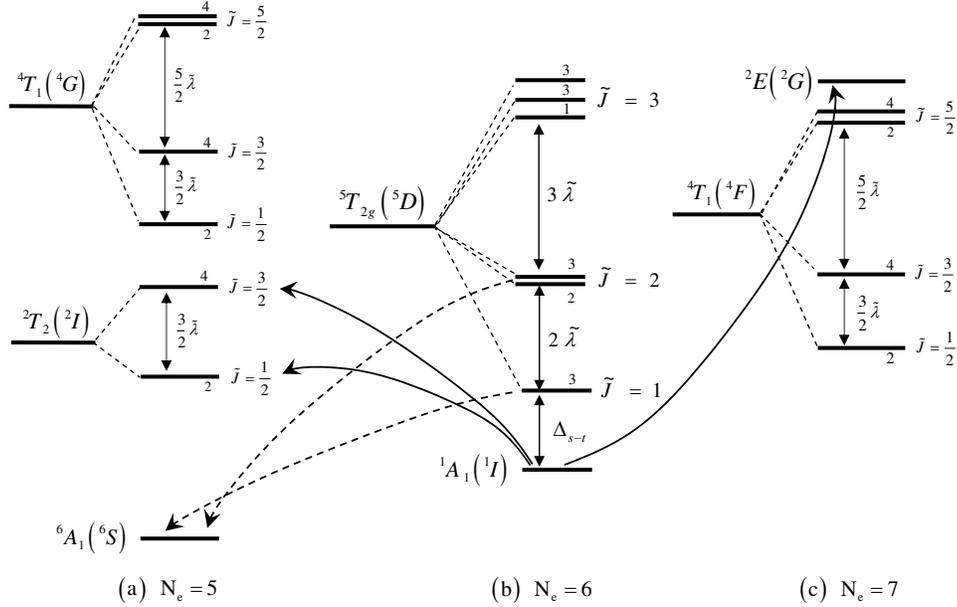}
 \caption{The low-energy part of the Hilbert space for CoO$_6$ cluster with the electron numbers $N_e=5, 6, 7$. Terms with a given $N_e$ are the mixtures of $d^{N_e}$, $d^{N_e+1}\underline{L}$, and $d^{N_e+2}\underline{L}^2$ configurations. At $T=0$, only the $N_e=6$ low-spin term $^1A_1$ is occupied; the Fermi-type excitations from this term which form the top of the valence band ($d^6 \to d^5$) and the bottom of the conductivity band ($d^6 \to d^7$) are shown by the solid lines with arrows. The dashed lines denote the in-gap excitations with the spectral weight increasing with temperature due to the population the HS excited $d^6$ terms.}
 \label{fig7}
\end{center}
\end{figure}
\begin{figure}[t]
\begin{center}
 \includegraphics*[width=1.0\textwidth]{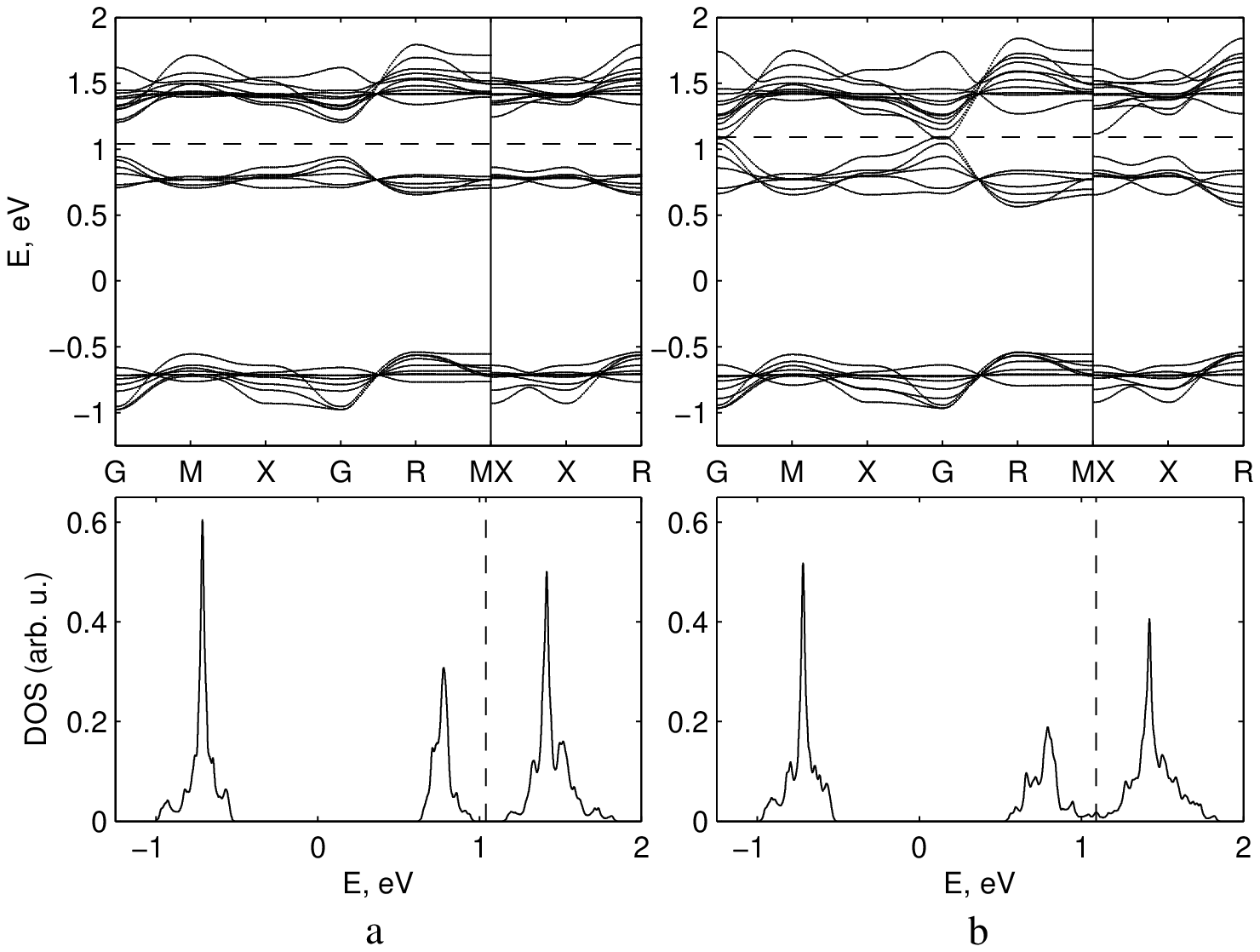}
 \caption{Quasiparticle dispersion and density of states at two different temperatures, $T = 100$K (a) and $T = 600$K (b). At $T = 0$K, LaCoO$_{3}$ is the charge-transfer insulator with the gap $E_{g} \approx 1.5$eV. At finite temperatures, the in-gap band appears below the conductivity band with the temperature dependent activation energy. At $T = 100$K, $E_{a} \approx 0.1$eV (a). At $T = T_\mathrm{IMT} = 587$K, $E_{a} = 0$eV, and above the $T_\mathrm{IMT}$ the band structure is of the metallic type (b).}
 \label{fig8}
\end{center}
\end{figure}

LaCoO$_{3}$ as well as other strongly correlated oxides is a difficult
problem for the ab initio band theory. The LDA calculations\cite{orlov11} incorrectly predict a metal for paramagnetic LaCoO$_{3}$. Various methods have been applied to study effect of correlations on the LaCoO$_{3}$ electronic
structure: LDA+U or GGA+U\cite{orlov12,orlov13,orlov14,orlov15}, dynamical mean-field theory\cite{orlov16}. Recent variational cluster approximation (VCA) calculation\cite{orlov17} based on the exact diagonalization of the CoO$_{6}$ cluster gives a reasonably accurate description of the low temperature properties: the insulating nature of the material, the photoelectron spectra, the LS-HS spin-state transition. The main deficiency of the VCA is the failure to reproduce the high temperature anomalies in the magnetic and electronic properties associated with the IMT.

The exact diagonalization of the multielectron Hamiltonian for a finite
cluster provides a reliable general overview of the electronic structure of
the correlated materials\cite{orlov18}. To incorporate the lattice effect several versions of the cluster perturbation theory are known\cite{orlov19,orlov20}. To calculate the band dispersion in the strongly correlated material one has to go beyond the local multielectron language. The natural tool to solve this problem is given by the Hubbard $X$-operators $X_f^{pq} = \left| p \right\rangle \left\langle q \right|$ constructed with the CoO$_{6}$ cluster eigenvectors $\left| p \right\rangle$ at site $\vec {R}_f $. All effects of the strong Coulomb interaction, spin-orbit coupling, covalence and the crystal field inside the CoO$_{6}$ cluster are included in the set of the local eigenstates $E_{p}$. Here $p$ denotes the following quantum numbers: the number of electrons (both $3d$ Co and $p$ of O), spin $S$ and pseudoorbital moment $l$ (or the total pseudomoment $J$ due to spin-orbit coupling), the irreducible representation in the crystal field. A relevant number of electrons is determined from the electroneutrality, for stoichiometric LaCoO$_{3}$, $n = 6$. In the pure ionic model the corresponding energy level scheme for $d^{6}$ Co$^{3+}$ has been obtained in\cite{orlov8}. Due to the covalence there is admixture of the ligand hole configurations $d^{n+1}{L}$ and $d^{n + 2}{L}^2$ that is very well known in the X-ray spectroscopy\cite{orlov21}. Contrary to spectroscopy the electronic structure calculations require the electron addition and removal excitations. For LaCoO$_{3}$ it means the $d^{5}$ and $d^{7}$ configurations. The total low energy Hilbert space is shown in the Fig.~\ref{fig7}. Here the energy level notations are the same as in the ionic model\cite{orlov8} but all eigenstate contains the oxygen hole admixture due to the covalence effect. The calculation of the $n = 5, 6, 7$ eigenvectors for CoO$_{6}$ cluster with the spin-orbit coupling and the Coulomb interaction has been done in\cite{orlov22}.

The electron removal spectrum determines the top of the valence band, the
corresponding electron quasiparticle QP are shown in the Fig.~\ref{fig7} by thin
solid lines as the excitation from the $^{1}A_{1}$ $d^{6}$ singlet in the
$^{2}T_{2}$ $d^{5}$ states with $J = 1/2$ and $J = 3/2$.
There energies are
\begin{eqnarray}
\Omega_{V1} &=& E \left( {d^6, \; ^1A_1 } \right) - E\left( {d^5, \; ^2T_2, \; \tilde {J} = 1/2} \right), \\
\Omega_{V2} &=& E\left( {d^6, \; ^1A_1 } \right) - E\left( {d^5, \; ^2T_2, \; \tilde {J} = 3/2} \right).
\end{eqnarray}

The bottom of empty conductivity band has the energy
\begin{equation}
\Omega_C = E\left( {d^7, \; ^2E} \right) - E\left( {d^6, \; ^1A_1} \right).
\end{equation}
All these bands have nonzero QP spectral weight. The intercluster hopping
results in the dispersion, $\Omega_n \to \Omega_n(k)$. The QP LDA+GTB
band structure corresponds to the insulator with the gap $E_{g} \approx 1.5$eV (Fig.~\ref{fig8}) at $T = 0$. This gap value is rather close to the VCA gap\cite{orlov17} and the experimental value $E_{g} \approx 1$eV\cite{orlov10}.

At finite temperature, the thermal excitation over the spin-gap $\Delta_{S}$ into the $J = 1$ and over the gap $\Delta_{S} + 2\lambda$ into the $J = 2$ sublevels of the HS $^{5}T_{2g}$ state occurs. We take $\Delta_{S} = 140$K and $\lambda = -185$K following Ref.~\refcite{orlov6}. Partial occupation of the excited HS states results in the drastically change of the QP spectrum. For $T = 0$K, excitations from the $^{1}A_{1}$ $d^{6}$ singlet in the lowest
$^{6}A_{1}$ $d^{5}$ term were forbidden due to spin conservation (the
corresponding matrix element $\gamma_{n} = 0$), and the excitation from
$\left| {d^6, \; \tilde {J} = 1} \right\rangle $ in $\left| {d^5, \; ^6A_1 }
\right\rangle $ has nonzero matrix element (shown by dashed line $\Omega
_{V1}^*$ in Fig.~\ref{fig7}) but zero filling factor as the excitation between two empty states. For $T \ne 0$, the filling factor for the $\Omega_{V1}^*$ and $\Omega_{V2}^*$ QP is nonzero and is equal to the occupation number $n_{1}$ and $n_{2}$ of the states $\left| {d^6, \; \tilde {J} = 1} \right\rangle $ and $\left| {d^6, \; \tilde {J} = 2} \right\rangle $ correspondingly. The energies of these QP are
\begin{eqnarray}
\Omega_{V1}^* &=& E\left( {d^6, \; ^5T_{2g}, \; \tilde {J} = 1} \right) -
E\left( {d^5, \; ^6A_1} \right), \\
\Omega_{V2}^* &=& E\left( {d^6, \; ^5T_{2g}, \; \tilde {J} = 2} \right) -
E\left( {d^5, \; ^6A_1} \right).
\end{eqnarray}

Energies of these QP appear to be slightly below the bottom of the
conductivity band, see DOS at finite temperature in Fig.~\ref{fig8}. Thus we have
obtained that these temperature-induced QP states lies inside the
charge-transfer gap, they are the in-gap states. Similar in-gap states are
known to results from doping in the high temperature superconductors. The
LaCoO$_{3}$ is unique because the in-gap states are induced by heating. The
chemical potential lies in the narrow gap $2E_{a} \quad  \approx 0.2$eV at $T = 100$K between the in-gap states and conductivity band.

From the GTB dispersion equation (\ref{PTstep:eq27}) it is clear that the in-gap bandwidth is proportional to the occupation numbers $n_{1}$ and $n_{2}$ of the excited HS states. With further temperature increase, the in-gap bands $\Omega_{V1}^*$ and $\Omega_{V2}^*$ become wider and finally overlap with the conductivity band $\Omega_C$ (Fig.~\ref{fig8}) at $T = T_\mathrm{IMT} = 587$K. It should be clarified that the IMT in LaCoO$_{3}$ is not the thermodynamic phase transition, there is no any order parameter associated with the gap contrary to the classical IMT in VO$_{2}$, NiS \textit{etc}.

Thus, we find that a correct definition of the electron in strongly
correlated system directly results in the in-gap states during the
spin-state transition due to the thermal population of the excited HS
states. Close to the spin-state temperature region the in-gap states
determine the value of the activation energy $E_{a} \approx 0.1$eV.
Further temperature increase results in large in-gap bandwidth and smaller
$E_{a}$, and finally $E_{a} = 0$ at $T_\mathrm{IMT}$. As concerns the weak
maximum in the $\chi(T)$ close to the IMT, it may be a small Pauli-type
contribution from the itinerant carriers above $T_\mathrm{IMT}$. We emphasize that instead of rather large difference in temperatures of the spin-state
transition ($\sim 100$K) and the IMT (600K) the underlying mechanism is the
same and is induced by the thermal population of the excited HS states.

\section{Conclusions}
\label{EndGTBLDA:8}
We have presented the main ideas of the LDA+GTB method. Being invented to study the high-$T_c$ superconductivity in cuprates, LDA+GTB method appears to be powerful approach to systems with SEC and useful for other Mott insulators. Since it is a combination of the \textit{ab initio} and model approaches, the method cannot go beyond the restriction of the model used in GTB. For example, the absence of the long-range Coulomb interaction which determines the Coulomb matrix elements in the large-wavelength limit prevents the correct description of the overdoped cuprates. Of course, it is the common deficiency of all Hubbard-type models. The modern version of the LDA+GTB cannot be used when the perturbation parameter $t/U$ increases and the Mott transition is expected. Nevertheless, it works in the most difficult for conventional band theory region of strong electronic correlations.

Application of the LDA+GTb method for cuprates revealed \textit{two} critical points in the doping dependence. The first one is related to the change of the FS connectivity and logarithmic divergences of DOS and of electronic heat capacity parameter $\gamma$ at the optimal doping $p_{opt} = 0.151$. The logarithmic enhancement of DOS leads to the maximum in the doping dependence of superconducting critical temperature $T_c$ at the same critical point $x = p_{opt}$. The second QPT is associated with the collapse of the electron-like FS pocket at $p \to p^* = 0.246$ and results in the step singularities in DOS and in Sommerfeld parameter $\gamma$. We have found the depletion of the hole's kinetic energy below $p^*$ and ascribe it to the pseudogap formation at $p <
p^*$. Thus the two energy scales in cuprates measured by $T_c$ and $T^*$ are
both related to the QPTs and to the changes of the cuprate's electronic structure with doping. The underlying physics is tightly connected with the scattering on the spin fluctuations. While approach starting from the conventional Feynman diagrammatic expansion requires frequency-dependent self-energy to get the same results (see, e.g., CDMFT), $X$-operators technique allows to catch the crucial effects of scattering even on the mean-filed level.

The multiorbital extension of the same ideas results in a qualitatively correct description of the main peculiarities of the electronic properties of manganites and cobaltites. Further study of these and other Mott-Hubbard insulators are in progress.

\section*{Acknowledgments}

We are thankful to O.K. Andersen, V.I. Anisimov, A.F. Barabanov, K.I. Kikoin, N.M. Plakida, S. Sakai, A.-M.S. Tremblay, V.V. Val'kov, and R.O. Zaitsev for useful discussions.

This work was supported by the Presisium of RAS program ''Quantum physics of condensed matter'' N~20.7, Grant ``Leading scientific schools of Russia'' (NSh 1044-2012.2), RFBR (Grant No. 09-02-00127), Integration Grant of SBRAS-UrBRAS N~44, Grant of President of Russia MK-1683.2010.2, FCP Scientific and Research-and-Educational Personnel of Innovative Russia for 2009-2013 (GK 16.740.12.0731 and GK 16.740.11.0740), and Siberian Federal University (Theme N~F-11). M.M.K. and E.I.S. acknowledges support from The Dynasty Foundation and ICFPM.

\end{document}